\documentclass[referee,useAMS,usenatbib,usegraphicx]{mn2e}
\usepackage{amssymb}
\usepackage{times}

\title[Flaring HST-1 Knot in M~87 Jet]{Dynamics and High Energy Emission\\ of the Flaring HST-1 Knot in the M~87 Jet}
\author[\L . Stawarz et al.]{\L . Stawarz$^{1, \, 2, \, 3, \,}$\thanks{E-mail: Lukasz.Stawarz@mpi-hd.mpg.de}, F. Aharonian$^{1}$, J. Kataoka$^{4}$, \newauthor M. Ostrowski$^{3}$, A. Siemiginowska$^{5}$, and M. Sikora$^{6}$\\
$^1$Max-Planck-Institut f\"ur Kernphysik, Saupfercheckweg 1, 69117 Heidelberg, Germany\\
$^2$Landessternwarte Heidelberg, K\"onigstuhl, D-69117 Heidelberg, Germany\\
$^3$Obserwatorium Astronomiczne, Uniwersytet Jagiello\'nski, ul. Orla 171, 30-244 Krak\'ow, Poland\\
$^4$Department of Physics, Tokyo Institute of Technology, 2-12-1, Ohokayama, Meguro, Tokyo 152-8551, Japan\\
$^5$Harvard-Smithsonian Center for Astrophysics, 60 Garden Street, Cambridge, MA 02138, USA\\
$^6$Centrum Astronomiczne im. M. Kopernika, ul. Bartycka 18, 00-716 Warszawa, Poland}

\begin{document}

\date{}

\pagerange{\pageref{firstpage}--\pageref{lastpage}} \pubyear{2006}

\maketitle

\label{firstpage}

\begin{abstract}
Stimulated by recent observations of a giant radio-to-X-ray
synchrotron flare from HST-1, the innermost knot of the M~87 jet, as
well as by a detection of a very high energy $\gamma$-ray emission
from M~87, we investigated the dynamics and multiwavelength emission
of the HST-1 region. We study thermal pressure of the hot interstellar 
medium in M~87 and argue for a presence of a gaseous condensation in its 
central parts. We postulate that this additional feature is linked to 
the observed central stellar cusp of the elliptical host. Interaction of 
the jet with such a feature is likely to result in formation of a 
stationary converging/diverging reconfinement/reflected shock structure in 
the innermost parts of the M~87 jet. We show that for a realistic set of
the outflow parameters, a stationary and a flaring part of the HST-1
knot located $\sim 100$ pc away from the active center can be
associated with the decelerated portion of the jet matter placed
immediately downstream of the point where the reconfinement shock
reaches the jet axis. We discuss a possible scenario explaining a
broad-band brightening of the HST-1 region related to the variable
activity of the central core. In particular, we show that assuming a
previous epoch of the high central black hole activity
resulting in ejection of excess particles and photons down along the
jet, one may first expect a high-energy flare of HST-1 due to
inverse-Comptonisation of the nuclear radiation, followed after a few
years by an increase in the synchrotron continuum of this region. The
synchrotron flare itself could be accompanied by a subsequent
inverse-Compton brightening due to upscattering of the ambient (mostly
starlight) photons. If this is the case, then the recently observed
order-of-magnitude increase in the knot luminosity in all spectral
bands could be regarded as an unusual echo of the order-of-magnitude
outburst that had happened previously (and could be eventually
observed some $\sim 40$ years ago) in the highly relativistic
active core of the M~87 radio galaxy. We show that very high energy
$\gamma$-ray fluxes expected in a framework of the proposed scenario
are consistent with the observed ones. 
\end{abstract}


\section{Introduction}

A kiloparsec-scale jet in M~87, the giant elliptical radio galaxy in
Virgo cluster --- the very first extragalactic jet ever discovered
\citep{cur18} --- provides us with an exceptional laboratory for
studying physics of relativistic collimated outflows. This is because
its proximity ($16$ Mpc, 1 arcsec = 78 pc) allows for
observations of the jet and of ambient medium at different
frequencies, with a very high spatial resolution. This jet has in fact
been studied in great detail in all wavelengths. One of the most
remarkable features of this jet is the inner HST-1 knot region,
observed at about 60 pc from the active core
\citep{bir99,per03,har03,har06}. Complexity of this innermost part of the
outflow, which consists of stationary \emph{and} superluminal
sub-components, as well as its uniquely variable broad-band emission,
calls for an explanation. Here we investigate the properties of HST-1
knot and present an attempt to provide such an explanation.

Below we summarize observational constraints on the physics of the M~87
jet in general. Next, in section 2, we investigate properties of the 
ambient medium necessary for understanding the dynamics of the
jet. 
We argue that the enhanced thermal pressure connected with a stellar cusp in
the innermost parts of the host galaxy is likely to form a stationary
converging/diverging reconfinement/reflected shock structure in the
jet. In section 3, we conclude that indeed the stationary and flaring
upstream edge of HST-1 knot can be associated with the decelerated
portion of the jet matter placed immediately downstream from the point
where the reconfinement shock reaches the jet axis. The presented
interpretation gives us a framework in which one can model the
broad-band emission of this part of the jet, and its high energy
$\gamma$-ray emission in particular. In section 4, we present an
evaluation of the radiation fields at the position of HST-1 knot. We
find that the energy density of the starlight and of the nuclear
emission can be comparable to the energy density of the equipartition
magnetic field in this jet region. This means that the high energy
$\gamma$-ray inverse-Compton emission of HST-1 knot's flaring
point should be expected at the (very roughly) similar level as its
observed radio-to-X-ray synchrotron emission. Since the latter one has
recently increased significantly up to $\sim 10^{42}$ erg s$^{-1}$
\citep{har06}, the expected TeV flux from HST-1 knot should then be
promisingly comparable to the one detected from the M~87 system
\citep{aha03,bei05}. This issue is investigated further in section 5. 
There we present a possible scenario relating variable emission of
the HST-1 knot/reconfinement nozzle with a modulated activity of the
relativistic central core. In particular, we show that assuming the
previous epoch of the high nuclear activity of a central black hole,
resulting in ejection of excess particles and photons down along the
jet, one may expect first a high-energy flare of HST-1 due to
inverse-Compton scattering of the nuclear radiation, followed a few
years later by an increase in the synchrotron continuum emission of
this region. Interestingly, the predicted $\gamma$-ray fluxes
(assuming energy equipartition between radiating electrons and the
jet magnetic field) are consistent with the observed ones. 
Final summary and conclusions from our study are presented in the last section 6.

\subsection{M~87 Jet}

{\it Very Long Baseline Interferometry} observations reported by \citet{jun99} show a
presence of a very broad radio-emitting limb-brightened outflow close
($\sim 10^{-2}$ parsec) to the M~87 center, characterized by an opening
angle of $\Phi_{\rm obs} \sim 60\degr$. This outflow experiences a strong
collimation at the \emph{projected} distance equivalent to $\sim 100$
$r_{\rm g}$ (Schwarzschild radii) from the central supermassive black
hole (hereafter `SMBH'), i.e., with the appropriate conversion $r_{\rm
g} = 3.85$~$\mu$arcsec~$= 0.003$~pc, at the distance of $\sim
0.4$~mas $\approx 0.03$~pc. The collimation continues out to $\approx
10$~pc from the center, where the jet adopts an opening angle
$\Phi_{\rm obs}
\la 10\degr$ that remains roughly stable further away from
the core. A detection of synchrotron self-absorption features in the
radio spectrum of the M~87 nuclear region allowed the placement of an upper
limit on the jet magnetic field $B < 0.2$ G at $r \sim 0.06$ pc
\citep{rey96}. Recently, by modeling a turn-over frequency 
along the jet in the radio spectra,
\citet{dod05} found $B \sim 0.01 - 0.1$ G for $r < 3$ mas $\approx
0.25$ pc, and $B < 0.01$ G further away along the jet.

The bright components of the radio jet placed at $r \sim 0.1-5$ pc
from the jet base are characterized by no, or some sub-relativistic
proper motions, $\beta_{\rm app} \leq 0.04$
\citep{jun95,dod05}. Further out along the jet, the knots detected at
several pc from the core (e.g., the knot L placed at $r \sim 0.16''
\approx 12.5$ pc) are also sub-luminal, however with slightly larger
apparent velocities $\beta_{\rm app} \sim 0.3-0.6$
\citep{rei89,bir99}. Surprisingly, several knots placed even further
out ($0.8''-6.3''$) occurred to be highly superluminal, with
$\beta_{\rm app}$ reaching $6$. In particular, {\it Hubble Space 
Telescope} observations
reported by \citet{bir99} showed that the unresolved stationary
feature upstream of the HST-1 knot (at $0.8'' \approx 62$ pc) seems to
emit various components down the jet, both slow and fast, with the
maximum apparent velocity of $\beta_{\rm app} \sim 5-6$. Also, all the
components of knot D ($2.7''-4'' \approx 210-312$ pc) are
superluminal, with $\beta_{\rm app} \sim 2.5-5$. Finally, knot E
placed at $r \sim 6'' \approx 500$ pc from the core is characterized
by a relatively high velocity of $\beta_{\rm app} \sim 4$. All the
measured superluminal features, if interpreted as moving blobs,
suggest the bulk Lorentz factor for the $1''-6''$ portion of the jet
larger than $\Gamma \geq 6$, and the jet viewing angle less than
$\theta \leq 20\degr$ \citep{bir99}. (With $\theta \approx 20\degr$ all
the projected distances along the jet cited in this section should be
multiplied by a factor of $3$.)

A stationary feature placed at the upstream edge of HST-1 knot has
been flaring in the optical and X-rays since 2002. The results of {\it Very Large Array}, 
{\it Chandra X-ray Observatory} and {\it Hubble} monitoring programs presented by
\citet{har03} and \citet{per03} established month-to-year variability
of its radio-to-X-ray synchrotron continuum, with a comparable
amplitude over the entire broad waveband. The HST-1 knot is unresolved
by {\it Hubble}, indicating its spatial dimension smaller than $R \leq
0.02'' \approx 1.5$ pc. The equipartition magnetic field at the
position of this knot, when evaluated at the quiescence state and
neglecting corrections due to expected relativistic bulk velocity of
the radiating plasma, is of the order of $B_{\rm eq} \sim 10^{-3}$
G. Magnetic field lines thereby are predominantly perpendicular to the
jet axis, as suggested by polarization studies \citep{per03}. The
degree of the linear polarization decreases from $0.68$ upstream of
the HST-1 flaring region ($0.72'' \approx 56$ pc), to $0.46$ at the
position of the flux maximum ($0.8'' \approx 62$ pc), and then to
$0.23$ downstream of it ($0.92'' \approx 72$ pc). The most recent data
show that until the year 2005 the X-ray emission of HST-1 knot
increased by as much as a factor of $50$ \citep{har06}.

At the distance $r \sim 12.2'' \approx 950$~pc away from the center
the jet brightens significantly, forming a prominent knot (knot A)
followed by the subsequent knots (B and C), to disappear at
approximately $r \sim 2$~kpc into an amorphous radio lobe visible at
low radio frequencies \citep{owe00}. {\it VLA} studies indicate
subluminal apparent velocities of these outer jet components,
with $\beta_{\rm app}
\la 0.5-0.6$ \citep{bir95}. The kpc-scale jet, when observed in
radio, exhibits filamentary limb-brightened morphology
\citep{owe89}. Both its optical and X-ray emissions are synchrotron in
origin, similar to the inner parts of the outflow \citep{bir91,mei96,spa96,per99,per01,mar02,wil02}. 
All along the jet the radio-to-optical power-law slope is $0.65 \la \alpha_{\rm
R-O} \la 0.8$, while the optical-to-X-ray one $1.0 \la
\alpha_{\rm O-X} \la 1.9$ with the exception of HST-1 knot,
for which $\alpha_{\rm O-X} \approx 0.8-1.0$ \citep{per05,wat05}. This
indicates a general `broken power-law' character of the broad-band
synchrotron spectrum along the M~87 jet. As discussed in
\citet{sta05}, the kpc-scale jet's magnetic field is not likely to be lower
than $B_{\rm eq} \sim 300$ $\mu$G.

{\it HEGRA} Cherenkov Telescopes System detected the M~87 emission
with the photon flux of $F_{\gamma}(>0.73 \, {\rm TeV}) \approx 0.96
\times 10^{-12}$ cm$^{-2}$ s$^{-1}$ \citep{aha03}. Assuming a spectral
index for the observed emission $\alpha_{\gamma} = 2$, this
corresponds to the isotropic of luminosity $L_{\gamma}(0.73 \, {\rm TeV})
\approx 6.9 \times 10^{40}$ erg s$^{-1}$. The observations were taken
in the period 1998-99, when the HST-1 flaring region was in its
quiescence epoch. Different scenarios were proposed to account for the
detected TeV signal, including various versions of modeling M~87
active nucleus \citep[`misaligned' and `structured' BL
Lac;][]{bai01,rei04,ghi05,geo05}, but also a high energy emission of the
M~87 host galaxy \citep{pfr03} or of the kpc-scale jet \citep[of its
brightest knot A in particular;][]{sta03}. The evidence for
a year-timescale variability established by the subsequent {\it Whipple}
and {\it H.E.S.S.} observations \citep{leb04,bei05}, indicating a
likely decrease of the TeV signal from M~87 from 1999 till 2004 by
about an order of magnitude (down to $L_{\gamma}(0.73 \, {\rm TeV})
\approx 0.54 \times 10^{40}$ erg s$^{-1}$), excludes the later two
possibilities, imposing however interesting constraints on the kpc-scale jet
magnetic field intensity \citep{sta05}. At the same time,
\emph{synchrotron} radio-to-X-ray emission of the HST-1 flaring region
increased by more than an order of magnitude
\citep[see][]{har06}.

\section{Host Galaxy Emission Profiles}

{\it Chandra} studies presented by \citet{you02} demonstrate that the
X-ray surface brightness of Virgo~A cluster (centered at the
position of M~87 radio galaxy) is of the modified King profile
$\Sigma_{\rm X}(r) \propto [1 + (r/r_{\rm K})^2]^{-3 \, \beta + 0.5}$,
with the parameter $\beta = 0.4$ and the critical radius $r_{\rm K}
\approx 18''$. This implies a density profile of the X-ray emitting hot
gas $\rho_{\rm G}(r) \propto [1 + (r/r_{\rm K})^2]^{-3 \, \beta / 2}$,
i.e. $\propto r^{-1.2}$ for $r > r_{\rm K}$ \citep[see,
e.g.,][]{sar86}. Both the temperature and the abundance of this gas
decrease smoothly toward the cluster center, reaching at $r < 60''$
values of $k T_{\rm G} \la 1.5$ keV and $Z <0.5 \, Z_{\odot}$,
respectively \citep{boe01}. It is not clear if the abundance decrease
is real, or only apparent, caused by resonant line scattering
\citep[but see][]{gas02}. However, even with this ambiguity, one
can conclude from the X-ray observations that the number density of
hot thermal electrons in the M~87 host galaxy decreases from $\sim 0.15$
cm$^{-3}$ at $r \sim 30''$ to $\sim 0.03$ cm$^{-3}$ at $r \sim 100''$
from the active core \citep{you02,dim03}. All of these constraints are
in agreement with a general finding that the central electron number
density in giant ellipticals is typically $\sim 0.1$~cm$^{-3}$, and
declines as $\propto r^{-1.25}$ with the distance from their cores
\citep{mat03}. Unfortunately, even with the excellent spatial
resolution of {\it Chandra} the thermal gas X-ray emission profile
cannot be probed precisely in the innermost portions of M~87, $r <
10''$, because of numerous X-ray emission features there
\citep[see][]{fen04}. As argued below, one can instead use the optical
observations to constrain the parameters of this gaseous medium.

Optical observations of M~87 reported by \citet{you78} indicate that
the modified isothermal sphere model (usually applied to elliptical
galaxies) is inconsistent with the observed starlight emission profile
for the projected radii of $r < 10''$.  In particular, they showed a
presence of a central luminosity excess, explained by \citet{you78} in
terms of a dynamical effect of a SMBH placed at the center of the
galaxy on its stellar neighborhood. {\it Hubble} observations
\citep{lau92} confirmed the presence of this additional stellar
component in  agreement with the interpretation involving $M
\approx 3 \times 10^9 \, M_{\odot}$ SMBH strongly bounding nearby
stars and creating a central stellar cusp with an increased stellar
velocity dispersion \citep[in this context see
also][]{dre90,mac97}. The observed optical (starlight) surface
brightness profile of the host galaxy is therefore described by
$\Sigma_{\rm O}(r)
\propto r^{-b}$ with $b = 0.25$ for $r < 3''$ and $b = 1.3$ for $r >
10''$, and composed of two separate components, namely a central cusp
with the luminosity density $\ga 10^3 \, L_{\odot}$ pc$^{-3}$ (in
the $I$ filter for $r < 0.1''$), and a modified King profile with the
curvature radius $r_{\rm C} \approx 7''$ and a tidal radius $r_{\rm T}
\ga 10^2 \, r_{\rm C}$. The later component is consistent with a
general property of the elliptical galaxies, namely $\log r_{\rm T} /
r_{\rm C} \sim 2.2$ \citep{sil98}.

As discussed in \citet{you80}, in the case of the adiabatic growth of
a central SMBH (i.e., the growth at a rate slower than the dynamical
time scale of the stellar cluster but faster than the relaxation time
scale), the expected density profile of stars follows a power law,
$\rho_{\rm S}(r) \propto r^{-a}$ with $a = 1.5$ \citep[see
also][]{van99}. Hence, the expected starlight brightness profile for
the adiabatic stellar cusp is $\Sigma_{\rm O}(r) \propto r \times
\rho_{\rm S}(r) \propto r^{-0.5}$, i.e. much steeper than the one observed
in M~87. However, as noted by \citet{lau92}, the behavior discussed by
\citet{you78} is in fact an asymptotic one, holding for $r \rightarrow
0$, while in the outer regions of a cusp an expected profile should be
flatter, similar to $\propto r^{-0.25}$ observed in M~87. This central
stellar cusp profile joins smoothly with the galactic starlight
profile $\Sigma_{\rm O}(r) \propto r^{-1.3}$ at further distances from
the core, implying a stellar density dependence of $\rho_{\rm S}(r)
\propto r^{-2.3}$. 
Note that {\it Hubble} observations \citep{lau95,fab97} of elliptical galaxies show
that they never possess a homogeneous surface brightness 
distribution $\Sigma_{\rm O}(r)
\propto {\rm const}$ expected in the case of
a pure King-like profile, but they can be divided into two classes:
(i) the `core type' galaxies with the brightness profile described by
a broken power law with $\Sigma_{\rm O}(r) \propto r^{-b}$ and $b
\leq 0.3$ within some critical (break) radius $r < r_{\rm B}$ (in M~87
case $r_{\rm B} \approx 3''$), and (ii) the `power-law type' galaxies
characterized by a single value of $b \geq 0.5$ within the whole
central region. As noted recently by
\citet{der05}, radio loud galaxies are always of the `core type',
although not every `core type' galaxy is radio loud. Also, the break
radius is proportional to the galactic luminosity, roughly $r_{\rm B} / 
{\rm kpc} \sim L_{\rm V} / 10^{45} \, {\rm erg \, s^{-1}}$. One can
therefore suspect the presence of an additional thermal pressure
component within the host galaxies of radio-loud AGNs, `matching' the
central stellar cusps.

Our main assumption follows from the observational fact reported by
\citet{tri86}, that the optical and X-ray surface brightness profiles
for three bright Virgo~A ellipticals are almost identical, i.e. that
$\Sigma_{\rm O}(r) \propto \Sigma_{\rm X}(r)$. Since the starlight
emissivity is proportional to the number density of the stars, while
the X-ray (bremsstrahlung) emissivity is proportional to the square of
the hot gas number density, one obtains $\rho_{\rm S}(r)
\propto
\rho^2_{\rm G}(r)$. This result was considered by \citet{mat03} as a
general property of elliptical galaxies. Indeed, in the case of M~87
host galaxy, at the distances $r > 18''$ one observes $\Sigma_{\rm
O}(r) \propto r^{-1.3}$ leading to $\rho_{\rm S}(r) \propto r^{-2.3}$,
and at the same time $\Sigma_{\rm X}(r) \propto r^{-1.4}$ leading to
$\rho_{\rm G}(r) \propto [\Sigma_{\rm X}(r) / r]^{1/2} \propto
r^{-1.2}$ (see above). This is in a good agreement with the expected
behavior $\rho_{\rm S}(r) \propto \rho^2_{\rm G}(r)$. Therefore,
hereafter we also assume that in the inner parts of M~87 the
distribution of the hot thermal gas (i.e. of the pressure) follows
closely the distribution of the stars (i.e. of the mass). We also
assume for simplicity a constant temperature of the hot gas $k T_{\rm
G} \sim 1$ keV within $r < 60''$ \citep[see][]{dim03}, leading to the
pressure profile simply proportional to the gas density profile,
$p_{\rm G}(r) \propto \rho_{\rm G}(r)$, i.e.
\begin{eqnarray}
p_{\rm G}(r) = p_0 \times \left\{ \begin{array}{lll} \left({r \over r_{\rm B}}\right)^{-0.6} & {\rm for} & r < r_{\rm B} \\
 \left[1 +\left({r \over r_{\rm K}}\right)^2\right]^{-0.6} & {\rm for} & r \geq r_{\rm B} \end{array} \right. ,
\end{eqnarray}
\noindent
with the normalization $p_0 = 1.5 \times 10^{-9}$ dyn cm$^{-2}$
\citep{you02}. Here $r_{\rm B} = 3'' \approx 234$ pc, and $r_{\rm K} =
18'' \approx 1.4$ kpc. The
resulting distribution of the gas pressure is shown in Figure 1. One
can see that the gas pressure decreases from $\sim 10^{-8}$ dyn
cm$^{-2}$ at $r \sim 10$ pc up to $\sim 10^{-10}$ dyn cm$^{-2}$ at $r
\sim 10$ kpc. In addition, the pressure profiles adopted by \citet{fal85}
and \citet{owe89} in their studies of the M~87 jet are shown for comparison. 
The former one is $p(r) = 1.2 \times
10^{-9} \, (1 + (r/r_{\rm B})^4)^{-1/4}$ dyn cm$^{-2}$, while the
latter one is $p(r) = 5.1 \times 10^{-10} \, (r/ {\rm kpc})^{-0.35}$
dyn cm$^{-2}$ between $0.7$ and $2$ kpc, and $p(r) = 7.3 \times
10^{-10} \, (r/ {\rm kpc})^{-0.85}$ dyn cm$^{-2}$ for $r > 2$
kpc. Note that these two approximations imply a lower gas pressure
than the one adopted by us.

\begin{figure}
\includegraphics[scale=1.5]{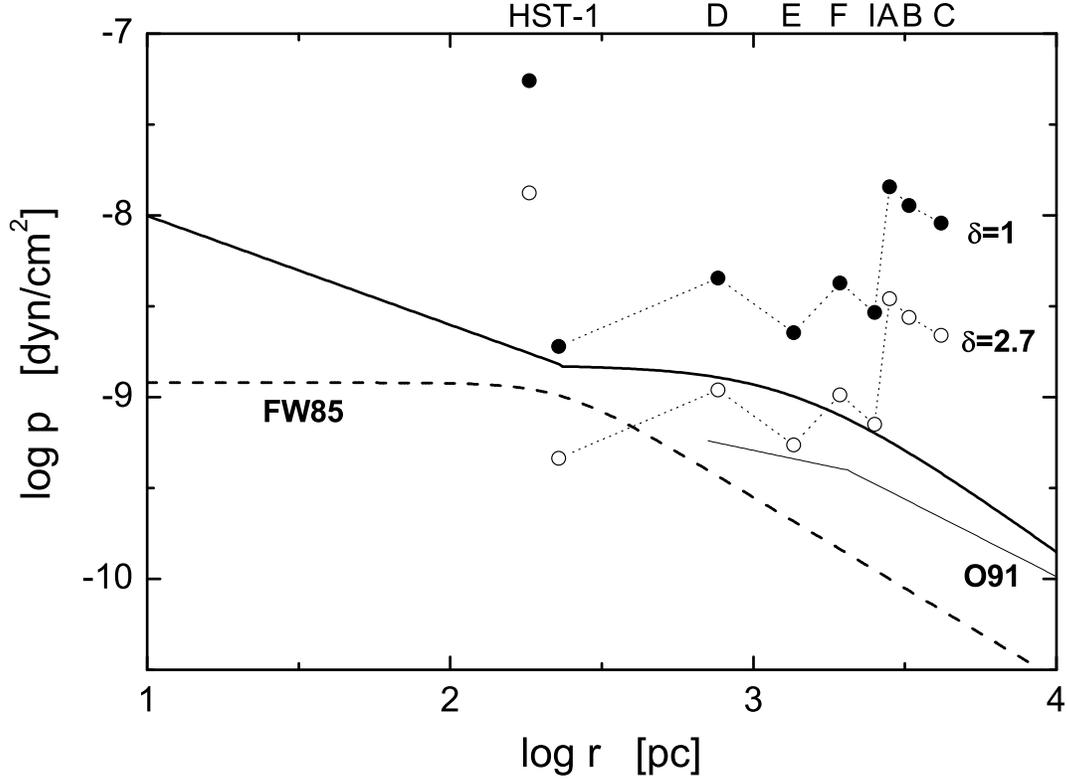}
\caption{Profiles of the hot gas pressure in M~87 host galaxy, as evaluated by 
\citet[dashed line]{fal85}, \citet[thin solid line]{owe89}, and in this paper 
(thick solid line). Circles indicate minimum pressure of the knots in the 
M~87 jet neglecting the relativistic correction (filled ones), 
and assuming the jet Doppler factor $\delta = 2.7$ (open ones). 
The circles disconnected from the others correspond to the HST-1 
flaring region (the upstream edge of the HST-1 knot). 
In deprojecting distances between the knots and the active core 
we assumed the jet viewing angle of $\theta = 20\degr$.}
\end{figure}

Figure 1 shows the \emph{de-projected} positions of different knots
(starting from HST-1 up to C) of the M~87 jet\footnote{Knot HST-1:
$0.8''-1.2''$, knot D: $2.7''-4''$, knot E: $5.7''-6.2''$, knot F:
$8.1''-8.8''$, knot I: $10.5''-11.5''$, knot A: $12.2''-12.5''$, knot
B: $14.1''-14.5''$, knot C: $17.5''-19''$.} assuming 
the jet viewing angle $\theta = 20\degr$
\citep{bic96,hei97}, and the minimum pressure of these knots (a sum
of the pressure due to ultrarelativistic radiating electrons and due
to the tangled magnetic field),
\begin{equation}
p_{\rm min} = p_{\rm eq, \, e} + U_{\rm B, \, eq} \approx 3.4 \times 10^{-9} \left({f_{\rm R} \over 100 \, {\rm mJy}}\right)^{4/7} \left({R \over 0.3''}\right)^{-12/7} \delta^{-10/7} \quad {\rm dyn \, cm^{-2}}
\end{equation}
\noindent
\citep[see][and Appendix B]{kat05}. Here $f_{\rm R}$ is the observed radio flux 
of the knot at $15$ GHz, $R$ is its observed knot's radius (assuming
spherical geometry), and $\delta$ is the knot's Doppler factor. In
Figure 1 we consider two jet's Doppler factors: $\delta = 1$ and
$\delta = 2.7$. The later one is appropriate for the expected jet
viewing angle $\theta \sim 20\degr$ and the jet bulk Lorentz factor
$\Gamma \sim 3 - 5$ on kpc scales
\citep{bic96,hei97}. We also took $R=0.3''$ (except the HST-1
flaring region, i.e., the upstream edge of HST-1 knot disconnected
from the other knots in Figure 1, for which we assume $R=0.02''$) and
used the knots' $15$ GHz fluxes given by \citet{per01}\footnote{Knot
HST-1 (total): $35.64$ mJy, knot D: $161.54$ mJy, knot E: $48.05$ mJy,
knot F: $144.9$ mJy, knot I: $75.8$ mJy, knot A: $1218$ mJy, knot B:
$808.4$ mJy, knot C: $544.7$ mJy.}. For the flaring region of HST-1
knot we took the $15$ GHz flux of $3.8$ mJy, as given in \citet{har03}
for the quiescence epoch of this part of the jet. Note, that the first
bright knot HST-1 is placed very close to $r_{\rm B}$, i.e. the radius
where the change in the ambient pressure profile between the central
cusp and the unperturbed King-like distribution is expected to
occur. In addition, downstream of this region, for $r_{\rm B} < r <
r_{\rm K}$, the M~87 jet is overpressured in respect to the gaseous medium
by a factor of a few, and even by more than an order of magnitude at
the position of the brightest knot A further away. However, with the
beaming effects included, the minimum pressure of the knots D, E, F
and I is almost the same as the ambient medium pressure. Note also
that the HST-1 flaring region is highly overpressured.

An additional gaseous X-ray condensation in the center of M~87 host
galaxy, linked to the observed in optical stellar cusp, increases a
thermal pressure of the galactic medium (with respect to the `pure'
King-like profile) by as much as an order of magnitude at the distance
$\sim 10$ pc from the core. On the other hand, a small volume occupied
by this additional component implies only a small excess X-ray thermal
luminosity. Figure 2 shows this light increase in the X-ray surface
brightness profile.  We calculate the X-ray surface brightness of host
galaxy with and without this central component. Because the
bremsstrahlung emissivity is proportional to the square of the thermal
gas density, the appropriate surface brightness, as a function of the
\emph{projected} distance from the nucleus $r_{\rm p}$, is
\begin{equation}
\Sigma_{\rm X}(r_{\rm p}) \propto \int_0^{l_{\rm max}} \, p_{\rm G}^2\left(\sqrt{l^2 + r_{\rm p}^2}\right) \, dl \, ,
\end{equation}
\noindent
where $l_{\rm max} = \sqrt{r^2_{\rm T} - r_{\rm p}^2}$ and $r_{\rm T}
= 10^{2.1} \, r_{\rm C} \approx 68.7$ kpc \citep{lau92}. We assumed a
constant temperature of the gaseous medium within the galaxy, and took
the pressure profile as given by equation 1 with and without the
central cusp. As shown in Figure 2, the additional central component
increases only slightly the X-ray surface brightness, in particular by
a factor of $2-3$ within $r_{\rm p}
\leq 100$ pc.

\begin{figure}
\includegraphics[scale=1.5]{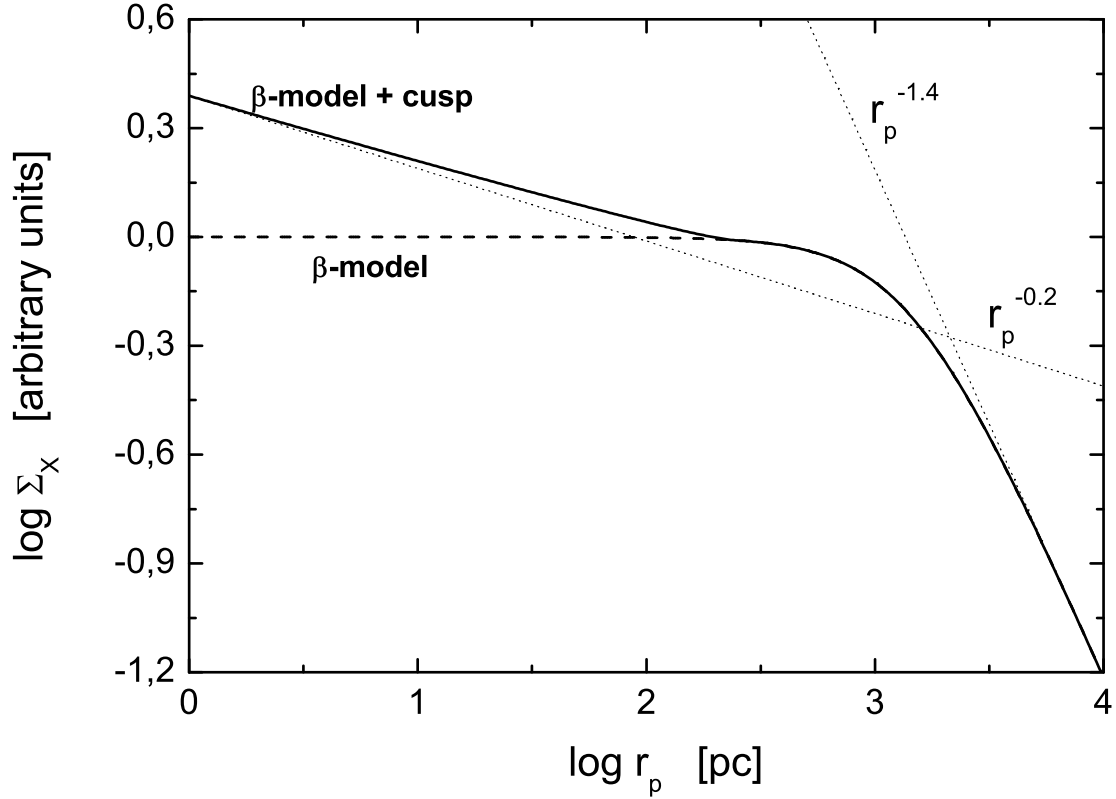}
\caption{Normalized X-ray surface brightness ($\Sigma_X$) profiles of the 
M~87 host galaxy due to emission of the hot gas. A dashed line
corresponds to the pure $\beta$-model for the hot gas distribution, a
solid line to the $\beta$-model with an additional contribution from the
central stellar cusp, while dotted lines indicate respective
power-law asymptotics.}
\end{figure}

Interestingly, \citet{hard02} reported larger, by a factor of two,
number of counts from a thermal X-ray halo surrounding the central
regions ($r_{\rm p} < 500$ pc) of FR I radio galaxy 3C~31 than the
number of counts expected from the pure $\beta$ model fitted to the
entire 3C~31 host galaxy profile. We believe that this excess can be
related to a condensation of the hot interstellar medium in the
central parts of the host galaxy, as discussed above. In the case of
the M~87 source, however, it would be difficult to claim a presence of
an analogous feature at $r < 100$ pc based on the available {\it
Chandra} data, due to extremely complicated M~87 X-ray structure 
consisting of gaseous rings, voids, as well as due to strong
non-thermal keV emission of the jet itself
\citep[see][]{fen04,dim03}. Because of such complexity, the thermal
pressure profile proposed above should be really considered as a
simple approximation only. For the purpose of the analysis presented
below, it is however accurate enough.

Let us note, that the spatial scale for the postulated here central gaseous 
condensation is very small when compared to the scale of the M~87 radio 
lobes. Thus, its presence does not contradict with widely discussed
disruption of cooling-flow atmospheres by kpc-scale radio outflows 
\citep[see, e.g.,][for the particular case of M~87 source]{bic96}. 
In fact, the sound-crossing time over the region with the spatial scale 
$\sim r_{\rm B}$ is less than $1$ Myr (for the interstellar 
medium parameters as considered in this section), i.e. less than the lifetime 
of the inner lobes in the M~87 radio galaxy \citep{bic96}, suggesting relatively
short timescale for formation/regeneration of the central gaseous cusp.

\section{HST-1 Knot as a Reconfinement Shock}

We do not intend to explain here the observed gradual collimation of
the M~87 jet in its innermost parts. Instead we note that the initial
collimation of the broad nuclear outflow may be due to a dynamically
dominating magnetic field \citep{gra05}, as the jets in active
galactic nuclei are most likely launched by the magnetohydrodynamical
processes. On the other hand, a dominant electromagnetic jet flux
should be converted at some point to the particle flux, since the
large-scale jets seem to be rather particle dominated \citep[see a
discussion in][]{sik05}. Let us therefore speculate, that at
sufficiently large distance from the nucleus --- where the initial
collimation is completed --- the relativistic jet in the  M~87 radio galaxy
is already particle dominated, and starts to expand freely. In such
freely expanding jet, the pressure decreases very rapidly with the
distance, $r$, from the core.  For example, in the case of cold jet
matter the thermal pressure goes as $p_{\rm j}(r) \propto r^{-2 \,
\hat{\gamma}} = r^{-10/3}$ for $\hat{\gamma} = 5/3$
\citep{san83}. At the same time, the ambient gas pressure decreases much
less rapidly: above we argue that in the M~87 galaxy one has $p_{\rm
G}(r) \propto r^{-\eta}$ with $\eta = 0.6$ for $r < 235$ pc. Hence, as
$\eta < 2$, accordingly to the discussion in \citet{kom97}, the
initially free jet in M~87 certainly (i) will become reconfined at
some point $r_0$, (ii) will develop a reconfinement shock at its
boundary, possibly leading to limb-brightenings of the reconfining
outflow, and moreover (iii) the converging reconfinement shock will reach the jet
axis at some further position along the jet, $r_{\rm cr}$, beyond
which the whole jet itself will come to a pressure equilibrium with
the external gas medium. A simple evaluation of the reconfinement shock
parameters is presented in Appendix A for the cases of the jet matter
described by a non-relativistic equation of state (hereafter `cold
jet'), as done previously in \citet{kom97} \citep[see also in this
context][]{san83,fal85,wil85,wil87,fal91,kom94}, and also for an
ultrarelativistic equation of state (hereafter `hot jet').

At what distance from the M~87 nucleus, $r_0$, does  the jet reconfinement
start? \citet{rei89} noted that at the projected distance $\sim
0.05'' \approx 4$ pc from the core the jet radio morphology (opening
angle, transverse intensity profile) changes. Further out, beyond
$\sim 0.1'' \approx 8$ pc, the jet brightness drops below the
detection level, and then increases again at $\sim 0.15'' \approx 10$
pc forming a weak radio knot L. Beyond this knot, the jet radio
brightness decreases again, until $\sim 0.8'' \approx 62$ pc where a
very bright knot, HST-1 appears (see section 1). Bearing in mind all
the difficulties and uncertainties present in measurements regarding
detailed morphology of the nuclear jet radio structure, we conclude
that it is reasonable to identify $r_0$ with the jet region between
$0.05''$ and $0.1''$ (i.e. $4-8$ pc) from the M~87 center, and to
assume that the jet thereby is already relativistic and particle
dominated. Indeed, the initial --- hydromagnetic by assumption ---
collimation of a broad nuclear outflow seems to be already completed at
smaller distances from the core. \citet{gia06} argue that the
Poynting-flux dominated nuclear outflows in AGNs become kinetic flux
dominated at distances $\ga 10^3 \, r_{\rm g}$, i.e., in the case
of M~87 radio galaxy, at about $\ga 0.1$ pc projected (for $\theta
= 20\degr$), in agreement with our assumption.

With $r_0 \sim 0.05'' - 0.1''$, one should expect the reconfinement
shock to reach the jet axis at $r_{\rm cr} \sim 3 \, r_0 \sim 0.15'' -
0.3''$ projected distance from the center in the case of a cold jet,
or at $r_{\rm cr} \sim 10 \, r_0 \sim 0.5'' - 1.0''$ in the case of a
hot jet (see Appendix A). In other words, if the jet at $r_0$ is
dynamically dominated by cold particles, $r_{\rm cr}$ is expected to
roughly coincide with the knot L, while for the ultrarelativistic jet
matter --- consisting of (mildly) relativistic particles plus magnetic
field --- $r_{\rm cr}$ should rather be identified with the HST-1
complex. Again, noting all the rough approximations used by us to
derive $r_0$ and $r_{\rm cr}$, below we argue that the latter
interpretation is more likely. When the reconfinement shock reaches
the jet axis, converging supersonic flow downstream of the
reconfinement shock is expected to create the second stationary
`reflected' shock. This reflected shock is in turn diverging from the
jet axis along the outflow, starting from $r_{\rm cr}$
\citep{kom97}. The jet pressure immediately beyond $r_{\rm cr}$ is
therefore expected to be higher than the ambient medium pressure. This
is qualitatively consistent with what is presented in Figure 1 for the
HST-1 complex. 
Therefore we postulate that the extremely compact and
overpressured HST-1 flaring point, present at the very beginning of
the HST-1 complex, is placed at $\ga r_{\rm cr}$ (thus favoring
hot jet scenario), while the outer parts of the HST-1 complex ---
superluminal features characterized by the minimum pressure 
in rough equilibrium with the surrounding medium (see Figure 1) --- 
can be identified with the region occupied by a diverging
reflected shock further away from $r_{\rm cr}$.

\begin{figure}
\includegraphics[scale=1.5]{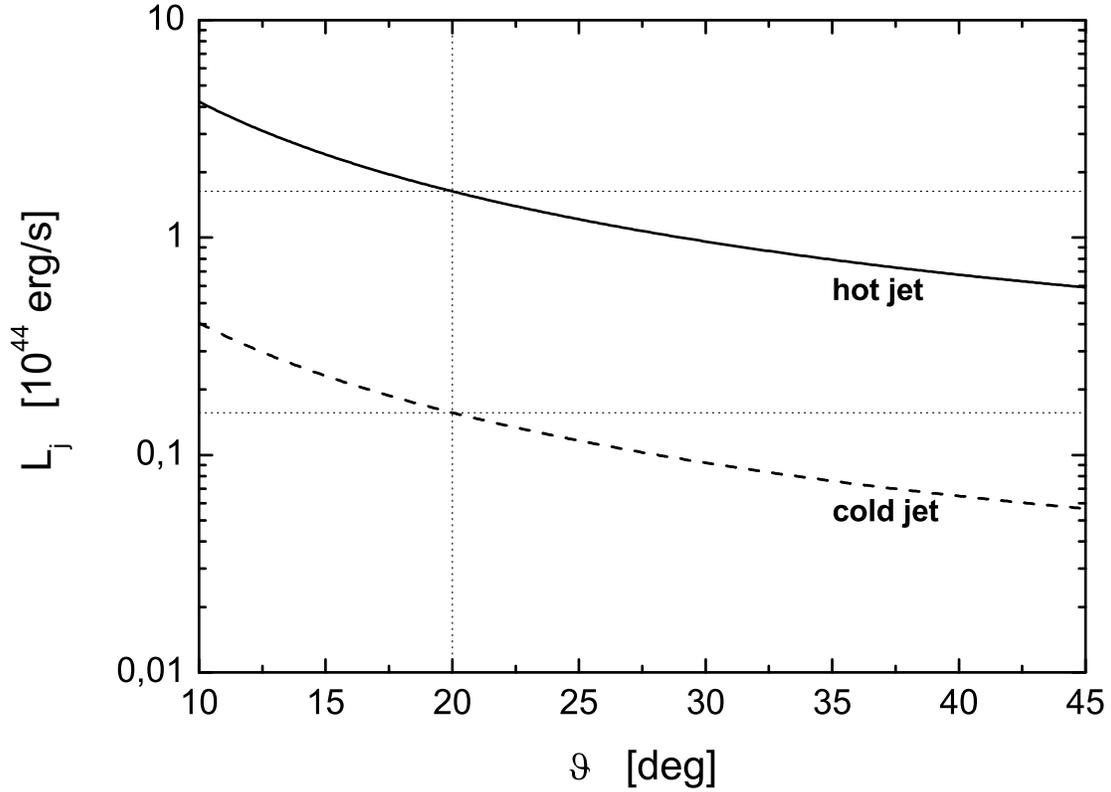}
\caption{A total kinetic power of the M~87 jet as a function of the 
jet viewing angle $\theta$, implied by the model in which the reconfinement shock
reaches the jet axis at the position of the HST-1 knot (solid line), 
and at the position of the knot L (dashed line).}
\end{figure}

For a given $r_{\rm cr} = r_{\rm cr, \, p} / \sin \theta$, where
$r_{\rm cr, \, p} = 0.8'' = 62.4$ pc is the projected distance of the
HST-1 flaring region and $\theta$ is the jet viewing angle, plus the
adopted ambient pressure profile $p_{\rm G}(r) \propto r^{-\eta}$ with
$\eta = 0.6$ and $p_0 = 1.5 \times 10^{-9}$ dyn cm$^{-2}$, kinetic
power of the jet implied by the model is
\begin{equation}
L_{\rm j} \sim 3 \, c \, \pi \, p_0 \, r_{\rm B}^{0.6} \, r_{\rm cr}^{1.4} \sim 0.4 \times 10^{44} \, \left(\sin \theta\right)^{-1.4} \quad {\rm erg \, s^{-1}}
\end{equation}
\noindent
(see Appendix A). The evaluated $L_{\rm j}$ is shown in Figure 3 for
different jet viewing angles. Note that for $\theta = 20\degr$ the
implied luminosity is $L_{\rm j} \approx 10^{44}$ erg s$^{-1}$,
consistent with the jet power required to feed radio lobes
\citep{bic96,owe00}. Figure 3 presents also the jet kinetic power
implied by the reconfinement shock position at $r_{\rm cr, \, p} = 0.15''
= 11.7$ pc (projected distance of the knot L), i.e. the location  preferred
in the cold jet scenario. In this case $L_{\rm j}$ is an order of
magnitude lower.

Previously, reconfinement shocks in FR I jets were proposed to be
placed at much larger distances from the central engines, namely at
the position of the brightest knots $\sim 1$ kpc from active nuclei
\citep{lai02}\footnote{For a possibility of stationary reconfinement
shocks in the small-scale jets of blazar sources see, e.g.,
\citet{jor01}.}. In the case of the M~87 jet it would be then at knot A
\citep{fal85}. In fact, it is possible that beyond HST-1 complex
($\ga 1.2''$) the M~87 jet breaks free again, and forms another
reconfinement shock around $\sim 10 \times 1.2'' = 12''$, i.e. exactly
at the position of knot A. On the other hand, knots beyond HST-1
complex were successfully explained by \citet{bic96} as oblique shocks
formed by helical modes of Kelvin-Helmholtz instabilities
characterized by a growing amplitude along the jet, disrupting finally
the outflow near knot C \citep[see also in this context][]{lob03}. For
our following analysis, the discussion on the dynamics of the
kpc-scale parts of the M~87 jet, beyond HST-1 is, however, not
crucial.

\section{Photon Fields}

We evaluate energy densities of the ambient radiation fields along the
jet axis, as measured in the rest frame of M~87 host galaxy at
different distances from the center. First, we note that the optical
starlight emission is dominated by photons at frequencies of $\sim
10^{14}$ Hz
\citep{mul04}, and that its emissivity profile is expected to follow
the galactic mass (i.e., star) distribution. Hence, the emissivity
is in a form
\begin{eqnarray} 
j_{\rm star}(r) = j_0 \times \left\{ \begin{array}{lll} \left({r \over r_{\rm B}}\right)^{-1.25} & {\rm for} & r < r_{\rm B} \\ 
\left[1 + \left({r \over r_{\rm C}}\right)^2\right]^{-1.15} & {\rm for} & r \geq r_{\rm B} \end{array} \right. ,
\end{eqnarray}
\noindent
with $j_{0} = 4 \times 10^{-22}$ erg s$^{-1}$ cm$^{-3}$, corresponding
to the $I$-band galaxy luminosity (see section 2). We integrate equation 5
along a ray and a solid angle with $r_{\rm T} = 10^{2.1} \, r_{\rm C}$
\citep[see][]{sta05} to obtain a profile of the starlight photons
energy density for M~87, $U_{\rm star}(r) = (1/c) \, \int j_{\rm
star}(r) \, ds \, d\Omega$, shown in Figure 4. Note that at distances
$r < 1$ kpc it is roughly constant with $\la 10^{-9}$ erg
cm$^{-3}$. We can also evaluate the energy density of the X-ray photons for 
the observed X-ray emission of the hot gas with the temperature $k \,
T_{\rm G} \approx 1.5$ keV in M~87.
We note that the bremsstrahlung emissivity is simply proportional to
the square of the gas number density, $j_{\rm ism}(r) \propto n_{\rm
G}^2(r)$, and hence, with the assumed constant gas temperature, to the
square of the gas pressure, $j_{\rm ism}(r) \propto p_{\rm
G}^2(r)$. By using the gas pressure profile given in the equation 1,
and integrating $j_{\rm ism}(r)$ along a ray and a solid angle with
the cluster termination radius $\sim 1$ Mpc, we obtain a distribution
of the X-ray photons energy density, $U_{\rm ism}(r)$, shown in Figure
4. Within the first kpc from the core the energy density of the
thermal X-ray photons is higher than the energy density of the cosmic
microwave background (CMB) photons, $\la 10^{-12}$ erg cm$^{-3}$.

Energy density of the diffused radiation from stars and hot
interstellar medium can be compared with the energy density of the
synchrotron emission produced within each knot of the M~87 jet. As
discussed in section 1, synchrotron emission of the knots 
is peaked at optical frequencies. Thus, in order to evaluate the
energy density of the synchrotron photons within the jet, we take the
optical fluxes measured at $10^{15}$ Hz by \citet{per01} for all the
knots\footnote{Knot D: $59.5$ $\mu$Jy, knot E: $16.2$ $\mu$Jy, knot F:
$62.7$ $\mu$Jy, knot I: $28.6$ $\mu$Jy, knot A: $586$ $\mu$Jy, knot B:
$306.8$ $\mu$Jy, knot C: $135.9$ $\mu$Jy.}, except of the HST-1
flaring point, for which we take $9$ $\mu$Jy, as given in
\citet{har03}, corresponding to its quiescence epoch. We also assume
a spherical geometry for the emission regions, with the radius $0.02''$ in
the case of the knot HST-1 (considering only its flaring component)
and $0.3''$ for the others. Figure 4 illustrates the resulting energy
density of the synchrotron photons along the M~87 jet (neglecting
relativistic corrections), $U_{\rm syn} = d_{\rm L}^2 \, [\nu_{\rm O}
f_{\rm O}] / (R^{2} \, c)$, where $f_{\rm O}$ is the optical flux of a
knot at $\nu_{\rm O} = 10^{15}$ Hz. In deprojecting distances of the 
knots from the active core jet viewing angle $\theta = 20\degr$ was
assumed for illustration.

\begin{figure}
\includegraphics[scale=1.5]{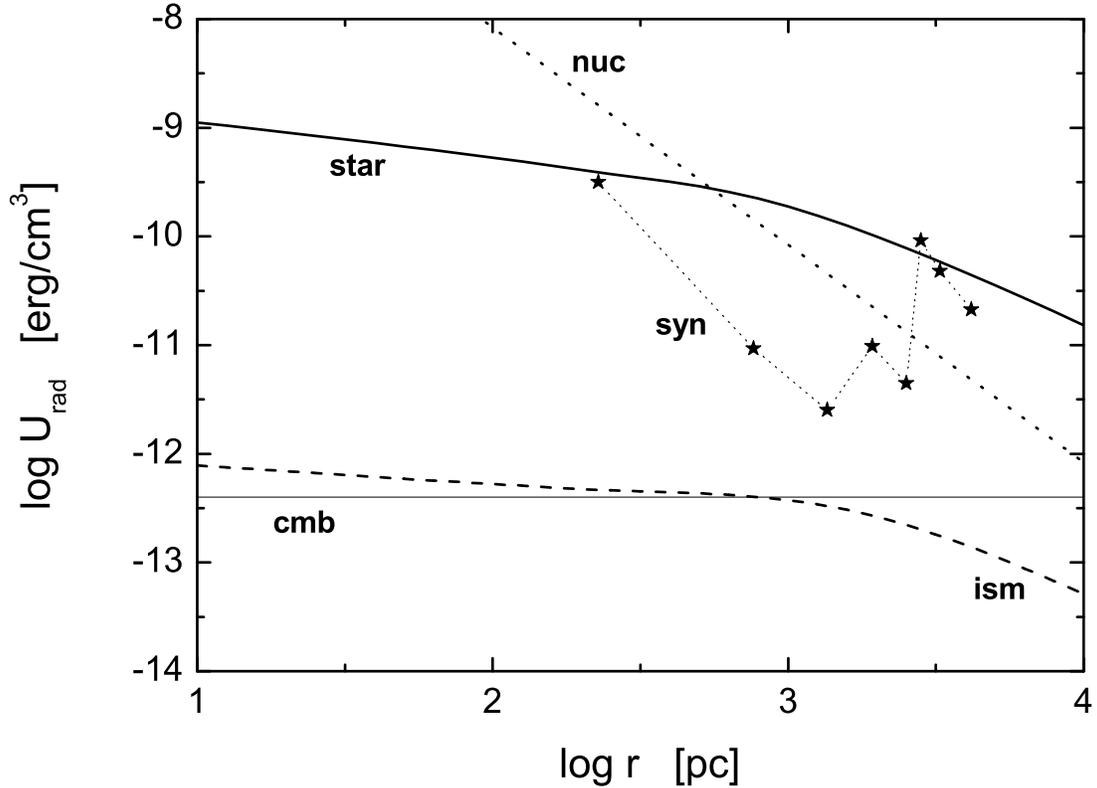}
\caption{Energy density profiles for different radiation fields 
as measured in the rest frame of the M~87 host galaxy 
by a stationary observer located at the jet axis. 
A thick solid line corresponds to the starlight emission, a 
dashed line to the thermal X-ray emission, a thin solid line to the CMB radiation, 
and a dotted line to the emission of the nuclear portion of the jet. 
Stars indicate energy densities of the internal synchrotron photons 
for different knots (neglecting relativistic corrections 
due to relativistic velocities of the emitting plasma).
In deprojecting distances between the knots and the active core 
we assumed the jet viewing angle of $\theta = 20\degr$.}
\end{figure}

Finally, any stationary observer located \emph{at the jet axis} is
illuminated by the radiation produced within the active
nucleus. Obviously, this emission is anisotropic, relativistically
beamed into a narrow cone depending on the (unknown) bulk Lorentz
factor of a nuclear jet. In this context we note, that the
sub-pc-scale ($r < 0.2$ mas $\approx 0.015$ pc) jet in M~87 has a
slightly different position angle than the large-scale jet in this
source, with the misalignment in position angle $\approx 15\degr$
\citep{jun95}. For these reasons, it is not obvious that the
large-scale jet is indeed illuminated from behind by the beamed
nuclear emission and what is the beaming amplification of such an
emission, i.e., if the jet flow at $r > 0.015$ pc from the center is
placed within the beaming cone of the nuclear jet. If, however, jet
misalignment can be neglected in this respect (because of relativistic
and projection effects which make apparent misalignment much larger
than the real one) then, as discussed in \citet{sta03}, energy density
of the nuclear jet emission in the galactic rest frame along the jet
axis is $U_{\rm nuc} = L_{\rm nuc} \, (2 \, \Gamma_{\rm nuc} /
\delta_{\rm nuc})^3 / (4 \pi \, r^2 \, c)$, where $L_{\rm nuc}$ is the
synchrotron luminosity of the nuclear jet observed at some viewing
angle $\theta$, $\Gamma_{\rm nuc}$ is the bulk Lorentz factor of the
nuclear jet and $\delta_{\rm nuc}$ is the appropriate nuclear Doppler
factor. For example, with $\theta \sim 20\degr$ and $\Gamma_{\rm nuc}
\sim 3 - 10$ one obtains $(2 \, \Gamma_{\rm nuc} / \delta_{\rm nuc})^3
\sim 10 - 10^3$. In Figure 4, for illustration we assume the nuclear
beaming correction factor $\sim 10^2$, and note that the uncertainty
in this approximation (for a fixed $\theta$) can be more than $\pm$
one order of magnitude. We further take $L_{\rm nuc} = 3 \times
10^{42}$ erg s$^{-1}$ \citep{tsv98} characterizing steady state of the
M~87 nucleus, obtaining thus a profile of $U_{\rm nuc}(r)$ shown in
Figure 4. Let us mention, that $L_{\rm nuc}$ is peaked at the observed
photon frequencies $\sim 10^{14}-10^{15}$ Hz.

In the rest frame of the jet, the energy densities of different
radiation fields depend on the bulk Lorentz factor and inclination of
some particular part of the jet. For example, the energy density of
the starlight emission (as well as of the cluster and CMB photon
field) are amplified in a plasma rest frame accordingly to $\propto
\Gamma^2$. Relativistic corrections also decrease the comoving energy
density of the synchrotron radiation accordingly to $\propto
\delta^{-3}$ (as appropriate for a stationary shock feature). Finally,
the nuclear emission in the rest frame of the outer jet is decreased by a
factor $(2 \, \Gamma)^{-2}$ \citep{sta03}. Note, that even varying 
the jet viewing angle alone influences deprojected distances of 
the jet features
and therefore the energy densities of the galactic and nuclear
radiation fields. In a framework of our model, HST-1 flaring point
corresponds to a compact region just downstream of the
reconfinement/reflect shocks system. Estimation of the appropriate
bulk Lorentz factor of the radiating plasma is not trivial in this case,
because we need to consider the oblique shock geometry. In particular,
the jet matter downstream of the reconfinement and reflected shock
fronts may still be relativistic, depending on the distance from the
jet axis (see Appendix A).

\begin{figure}
\includegraphics[scale=1.5]{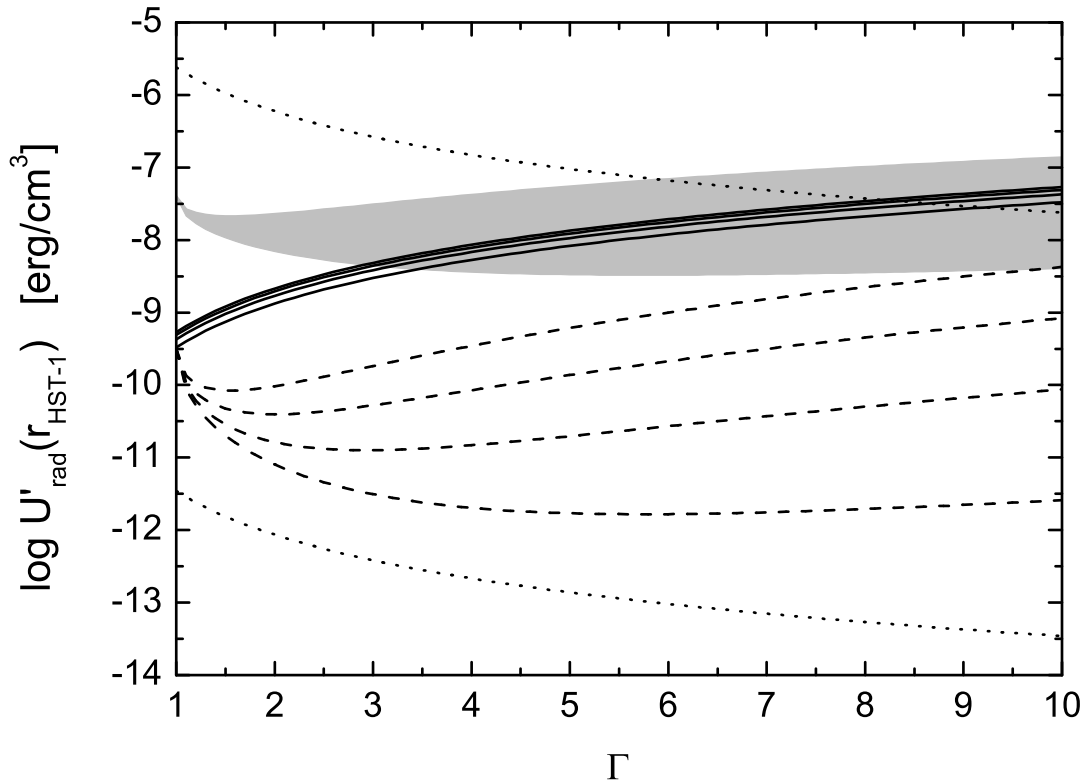}
\caption{Energy densities of different photon fields measured in a rest frame of the HST-1 flaring region, as functions of the bulk Lorentz factor of this part of the jet, $\Gamma$, for different jet viewing angles. Thick solid lines correspond to the starlight emission. Dashed lines correspond to the internal synchrotron emission of the knot, for the jet inclination $\theta = 10\degr$, $20\degr$, $30\degr$, and $40\degr$ (from bottom to top, respectively). Shaded region indicates energy density of the equipartition magnetic field for $\theta = 10\degr - 20\degr$. Dotted lines illustrate comoving energy density of the nuclear photons for $\theta = 40\degr$, $\Gamma_{\rm nuc} = 10$ (upper curve), and $\theta = 10\degr$, $\Gamma_{\rm nuc} = 3$ (lower curve).}
\end{figure}

If the HST-1 flaring point corresponds to a significantly decelerated
portion of the jet matter located at the very jet axis close to
$r_{\rm cr}$, the comoving energy densities of the starlight and
synchrotron photon fields are comparable, as presented in Figures 4
and 5 for different jet viewing angles. With an increasing bulk Lorenz
factor the energy density of the starlight emission increases from
about $\sim 3 \times 10^{-10}$ erg cm$^{-3}$ for $\Gamma \sim 1$ up to
$\sim 3 \times 10^{-8}$ erg cm$^{-3}$ for $\Gamma \sim 10$ (Figure
5). A shift in the deprojected position of the HST-1 flaring region
due to a different jet viewing angle is of negligible importance,
because the entire HST-1 complex is located within the central plateau
of $U_{\rm star}(r)$. The internal energy density of the jet synchrotron
emission initially decreases with a growing velocity of the
emitting region, but for the larger
values of $\Gamma$ it increases again, since for a given $\theta \geq
10\degr$ the appropriate jet Doppler factor $\delta$ decreases with an
increasing (large) $\Gamma$. One can however conclude, that for a wide
range of parameters shown in Figure 5 (namely, $\theta = 10\degr -
40\degr$ and $\Gamma = 1 - 10$), in the rest frame of the HST-1 flaring
region, the starlight emission is expected to dominate over the
internal synchrotron photon field. On the other hand, one should be
aware that as the HST-1 flaring point is unresolved (while at the
same time synchrotron energy density goes with the emission region
size as $\propto R^{-2}$), the estimated $U'_{\rm syn}$ should be
considered as a lower limit only. 
In addition, here we only considered a quiescent epoch of M~87. The
biggest uncertainties correspond however to a photon field of a nuclear jet
illuminating HST-1 knot from behind. Figure 5 illustrates two extreme
cases for $U'_{\rm nuc}$ at the position of this knot, corresponding
to the nuclear Lorentz factor $\Gamma_{\rm nuc} = 10$ and $\theta =
40\degr$, and also to $\Gamma_{\rm nuc} = 3$ and $\theta = 10\degr$. The
estimated energy density of the synchrotron emission of the nuclear
jet varies by a few orders of magnitude (!) for these two examples, and
may exceed or be much smaller than the other components, $U'_{\rm
star}$ and $U'_{\rm syn}$. Figure 5 shows also for a comparison the
energy density of the equipartition magnetic field, $U_{\rm B} =
(B_{\rm eq}^2 / 8 \pi)
\, \delta^{-10/7}$ with $B_{\rm eq} = 10^{-3}$ G, for $\theta = 10\degr
- 40\degr$.

\section{HST-1 Knot as a TeV Source?}

Let us suppose that the active core of M~87 experienced at some moment
an outburst, resulting in the flare of its synchrotron emission and
ejection of a portion of the ``jet matter'' with the excess kinetic
power (when compared to the steady-state epoch of the jet
activity). Both photons and particles travel along the jet, arriving
at some time to the location of the HST-1 knot, where the
reconfinement shock formed within a steady jet reaches the jet
axis. Flare synchrotron photons emitted by the active nucleus are then
comptonized to TeV energies (hereafter `IC/nuc' process), while the
excess jet matter shocked around $r_{\rm cr}$ causes synchrotron
(hereafter `SYN') and the additional inverse-Compton brightening of
the HST-1 flaring region. As discussed in the previous section and
also below, this additional inverse-Compton brightening should be
dominated by Compton scattering of the starlight emission (hereafter
`IC/star' process) or synchrotron self-Compton process (`SSC'). We
note, that some short time-scale variations of the emission coming
from the nucleus in its high state can be imprinted in the observed
IC/nuc flux of the outer parts of the jet. In addition, due to
different velocities of the nuclear photons and particles, TeV flare
resulting from the IC/nuc process in HST-1 knot is expected to lead
SYN, IC/star and SSC brightening of the HST-1 flaring region by some
time $\Delta t$. Note that the increase in the seed photon energy
density for a given synchrotron flux (i.e. for a given particle energy
density) results in an increase of the inverse-Compton flux only if
the electrons involved in the inverse-Compton scattering are weakly
cooled by radiative losses (`slow cooling regime').

Assuming that the observed sub-luminal velocities of the jet features
observed between the core and the HST-1 knot are only pattern
velocities, not reflecting the true bulk velocity of the jet spine
\citep[see a discussion in][]{dod05}, and that this true bulk velocity
is highly relativistic $\beta_{\rm nuc} \equiv (1 - \Gamma_{\rm
nuc}^{-2})^{-1/2} \sim 1$, the appropriate delay time difference is
roughly
\begin{equation}
\Delta t \approx {r \over c \, \beta_{\rm nuc}} - {r \over c} \approx {r_{\rm p}\over 2 c \, \Gamma_{\rm nuc}^2 \, \sin \theta} \sim 100 \, (\sin \theta)^{-1} \, \Gamma_{\rm nuc}^{-2} \quad {\rm yr} \, ,
\end{equation}
\noindent
where $r_{\rm p} = r \, \sin \theta = 62.4$ pc is a projected distance
of the HST-1 flaring region from the core. For example, period $\Delta
t \sim 6$ yr between presumable maximum of the TeV emission
(1998/1999) and the observed maximum of the synchrotron emission of
the HST-1 knot (2005) is consistent with the jet viewing angle $\theta
\sim 10\degr$ for $\Gamma_{\rm nuc} \sim 10$, with $\theta \sim 20\degr$
for $\Gamma_{\rm nuc} \sim 7$, and finally with $\theta \sim 30\degr$
for $\Gamma_{\rm nuc} \sim 6$. The assumed hypothetical nuclear flare
should be observed some $t_{\rm fl}$ years before the IC/nuc flare of
the HST-1 knot, where
\begin{equation}
t_{\rm fl} \approx {r_{\rm p} \over c} \, {1 - \cos \theta \over \sin \theta} \sim 200 \, (1 - \cos \theta) \, (\sin \theta)^{-1} \quad {\rm yr} \, .
\end{equation}
\noindent
For example, $t_{\rm fl} \sim 20$ yr for $\theta \sim 10\degr$, $t_{\rm
fl} \sim 35$ yr for $\theta \sim 20\degr$, and $t_{\rm fl} \sim 55$ yr
for $\theta \sim 30\degr$. We note in this context, that interestingly
\citet{dey71} reported radio flare of M~87 nucleus in 1969-1971. If
--- again for illustration --- one identifies the considered nuclear
flare with this event, then equations 6 and 7 imply $\theta \sim
16\degr$ and $\Gamma_{\rm nuc} \sim 8$.

\begin{figure}
\includegraphics[scale=1.5]{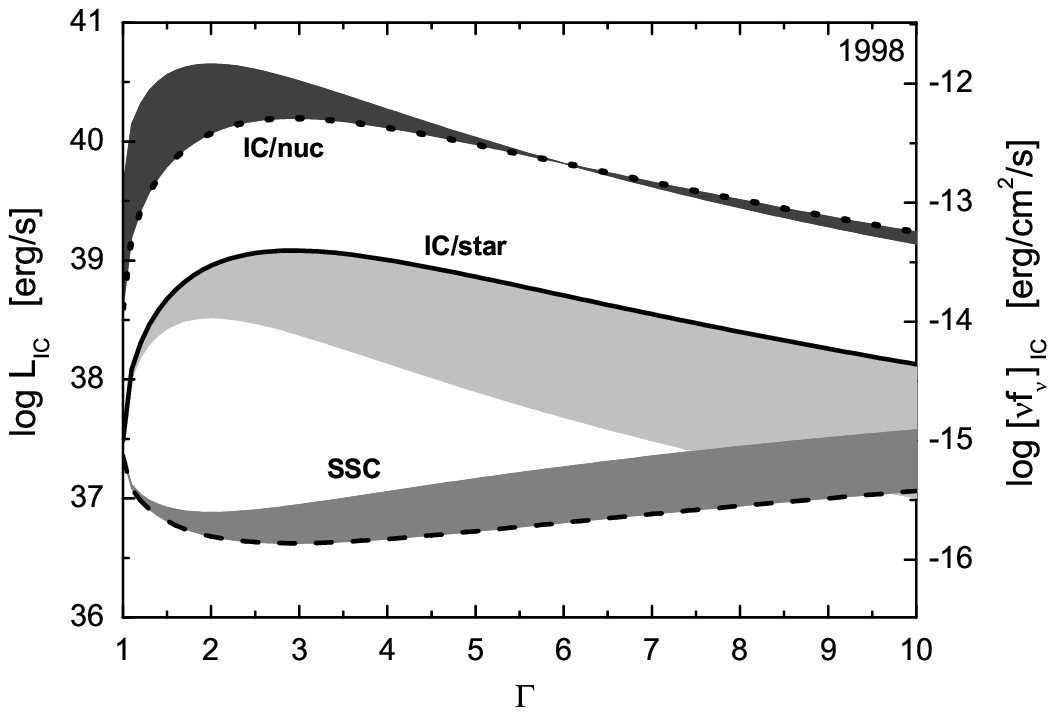}
\caption{Expected TeV emission of the HST-1 flaring region in 1998, 
due to IC/nuc (dotted line), IC/star (solid line), and SSC (dashed
lie) processes, as a function of the bulk Lorentz factor $\Gamma$ of
this part of the jet assuming $\theta = 20\degr$. Shaded regions
indicate the appropriate luminosity ranges for $\theta =
20\degr-30\degr$.}
\end{figure}

In the rest frame of the HST-1 knot, assuming moderate bulk velocity
and jet viewing angle, the energy densities of the starlight, nuclear and
internal synchrotron photons are peaked at similar photon frequencies
$10^{14} - 10^{15}$ Hz. Thus, electrons upscattering all these photons
to the observed TeV energies are mostly the slowly cooled ones, with
energies $\sim 10^6 \, m_{\rm e} c^2$ \citep[the optical spectral
index of HST-1 knot is consistent with $\alpha_{\rm O} \sim
0.6$;][]{per03}. The resulting TeV fluxes due to the IC/star, IC/nuc
and SSC processes 
are then produced in the transition between Thomson and Klein-Nishina
regimes. Hence, for a rough evaluations one can approximate the
expected observed TeV fluxes by $L_{\rm IC/seed} \sim f_{\rm \pm, \,
iso} \, (U'_{\rm seed} / U'_{\rm B}) \, L_{\rm O}$, where $L_{\rm O} =
4 \pi \, d_{\rm L} \, [\nu_{\rm O} f_{\rm O}]$ is the observed optical
luminosity of the HST-1 knot, while $U'_{\rm seed}$ and $U'_{\rm B}$
are the comoving energy densities of the appropriate seed photons and
the equipartition magnetic field. Function $f_{\rm \pm, \, iso} =
f_{\rm \pm, \, iso}(\Gamma, \theta)$ arises due to possible anisotropy
of the seed photons in the emitting region rest frame \citep[see][and
Appendix B]{sta03}. Using the observed radio fluxes $f_{\rm R}$ of the
HST-1 flaring region as measured at $15$ GHz, the optical flux $f_{\rm O}$
at $10^{15}$ Hz, and the emitting region size $R$ in arcseconds, one
obtains (see Appendix B)
\begin{equation}
L_{\rm IC/star} \sim 7.3 \times 10^{36} \, \left({f_{\rm O} \over {\rm \mu Jy}}\right) \left({f_{\rm R} \over {\rm mJy}}\right)^{-4/7} \left({R \over 0.02''}\right)^{12/7} \delta^{24/7} \quad {\rm erg \, s^{-1}} \, ,
\end{equation}
\noindent
\begin{equation}
L_{\rm SSC} \sim 0.6 \times 10^{36} \, \left({f_{\rm O} \over {\rm \mu Jy}}\right)^2 \left({f_{\rm R} \over {\rm mJy}}\right)^{-4/7} \left({R \over 0.02''}\right)^{-2/7} \delta^{-11/7} \quad {\rm erg \, s^{-1}} \, ,
\end{equation}
\noindent
and
\begin{eqnarray}
\lefteqn{ L_{\rm IC/nuc} \sim 7.4 \times 10^{39} \, \left({f_{\rm O} \over {\rm \mu Jy}}\right) \left({f_{\rm R} \over {\rm mJy}}\right)^{-4/7} \left({R \over 0.02''}\right)^{12/7} \left({\Gamma^2_{\rm nuc} L'_{\rm fl} \over 10^{45} \, {\rm erg/s}}\right) \, \delta^{24/7} {} } \nonumber\\
&& {} \times (\sin \theta)^2 \, (1 - \cos \theta)^2 \quad {\rm erg \, s^{-1}} \, .
\end{eqnarray}
\noindent
In the above we assumed that at every moment (i.e., for a given
$f_{\rm R}$ and $f_{\rm O}$) the HST-1 flaring region is in
equipartition regarding energies of the ultrarelativistic electrons
and the magnetic field. We also introduced internal nuclear luminosity
$L'_{\rm fl}$, which should correspond to the assumed nuclear
outburst, and not to the steady-state discussed in the previous
section. 

\begin{figure}
\includegraphics[scale=1.5]{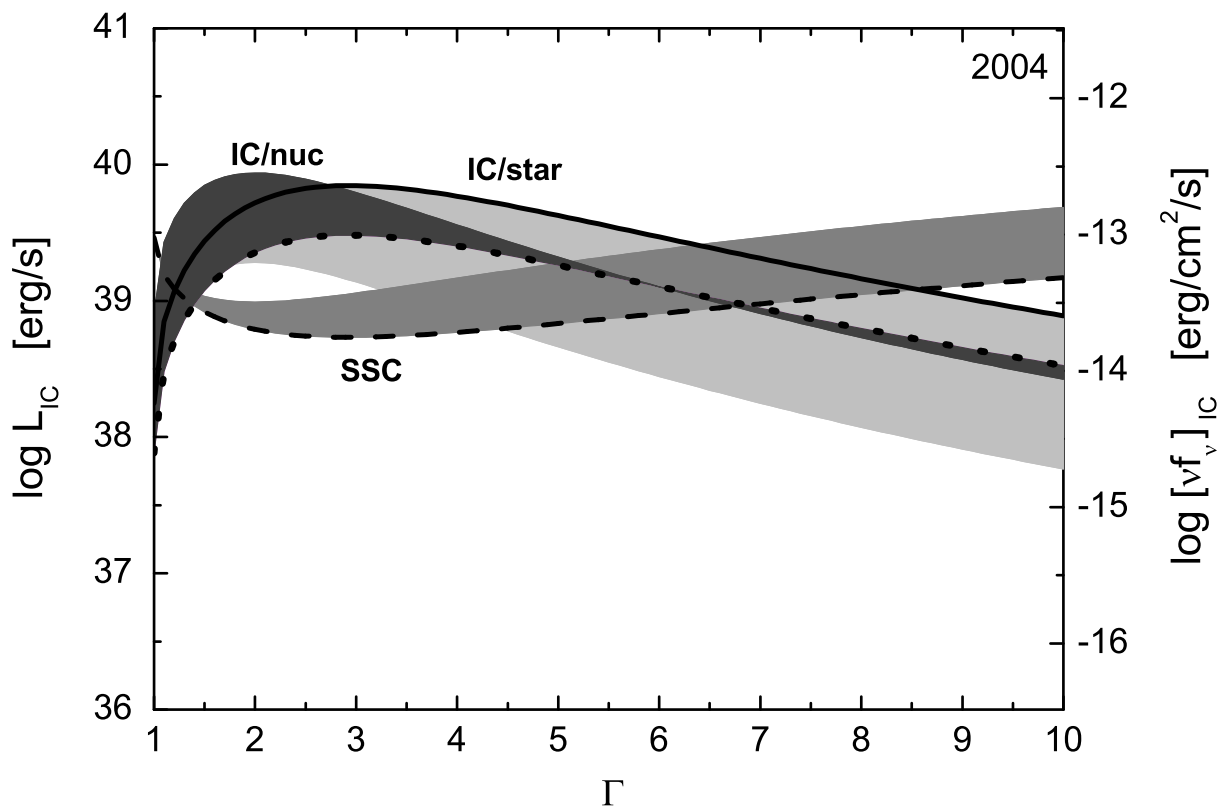}
\caption{Expected TeV emission of the HST-1 flaring region in 2004, 
due to IC/nuc (dotted line), IC/star (solid line) and SSC (dashed line) processes, 
as function of the bulk Lorentz factor $\Gamma$ of this part of the jet 
assuming $\theta = 20\degr$. Shaded regions indicate the appropriate luminosity 
ranges for $\theta = 20\degr-30\degr$.}
\end{figure}

Different constraints presented and cited in previous sections suggest
the most likely jet viewing angle of $\theta \sim 20\degr$. In Figure 6
we present the expected TeV emission of the HST-1 flaring region in
1998 assuming $\theta = 20\degr$, and resulting from IC/nuc, IC/star
and SSC processes (equations 8-10), as functions of the bulk Lorentz
factor of this part of the jet. For illustration, shaded regions
indicate also the appropriate luminosity expected for $\theta = 20\degr -
30\degr$. Here we took $f_{\rm R} = 3.8$ mJy, $f_{\rm O} = 9$ $\mu$Jy
and $R=0.02''$ \citep{har03}. The expected TeV IC/star emission is in this 
case $L_{\rm IC/star} < 10^{39}$ erg s$^{-1}$, and the corresponding SSC
emission is even lower. However, the IC/nuc emission could eventually
account for the {\it HEGRA} detection ($L_{\gamma} > 10^{40}$ erg
s$^{-1}$) only if the bulk Lorentz factor of the HST-1 flaring region was
$\Gamma \sim 2$ and the assumed nuclear flare was characterized by
$\Gamma_{\rm nuc}^2 L'_{\rm fl} \ga 3 \times 10^{46}$ erg
s$^{-1}$ (as taken in Figure 6). 
We note, that for such parameters the comoving energy density of the
nuclear photons, $U'_{\rm nuc} \ga 10^{-7}$ erg cm$^{-3}$,
dominates over the comoving energy densities of the magnetic field and
the starlight emission ($U'_{\rm B} \sim 10^{-8}$ erg cm$^{-3}$ and
$U'_{\rm star} \sim 10^{-9}$ erg cm$^{-3}$, respectively). Hence,
cooling of the considered $\sim$ TeV energy electrons is mainly due to
the IC/nuc process. We also note, that the obtained above value of the 
preferred bulk Lorentz factor $\Gamma \sim 2-3$ refers in our model 
strictly to the compact (unresolved) and decelerated portion of the outflow 
placed at the jet axis immediately after $r_{\rm cr}$, which is responsible 
for production of the flaring emission, and not to whole outflow at the same 
distance from the nucleus. In fact, oblique geometry of the reconfinement 
shock implies that the other parts of the jet, located further from the jet 
axis, may suffer much less deceleration, and thus that the average bulk Lorentz 
factor of the whole outflow may be higher than the one characterizing HST-1 
flaring region.

With the preferred $\theta \sim 20\degr - 30\degr$ and $\Gamma \sim
2-3$, the Doppler factor of the HST-1 flaring region is $\delta \sim
2-3$, while the jet-counterjet synchrotron brightness asymmetry is
$f_{\rm j} /f_{\rm cj} \sim [(1 + \beta \cos \theta)/(1 - \beta \cos
\theta)]^{2.5} \sim 10^2 - 10^3$ (still constistently with the
observational limits). Assuming $\Gamma_{\rm nuc} \sim 10$ \citep[in
agreement with values usually derived by means of modeling broad-band
emission of BL Lac objects;][]{urr95}, the Doppler factor of the
nuclear M~87 jet would be roughly $\delta_{\rm nuc} \sim 1$. This
implies that the nuclear outburst assumed in our model would be
observed with the isotropic luminosity $L_{\rm fl} \sim (\delta_{\rm
nuc} / \Gamma_{\rm nuc})^3 \, (\Gamma_{\rm nuc}^2 L'_{\rm fl}) \sim 3
\times 10^{43}$ erg s$^{-1}$. This is higher than the observed
luminosity of the M~87 nucleus in its steady-state epoch \citep{tsv98}
by only a factor $\sim 10$. We note, that order-of-magnitude flares on
time scales of years are common in blazar sources. In addition, as
mentioned above, synchrotron emission of the HST-1 flaring region has
increased between 1998 and 2005 by a similar factor $\sim 50$
\citep{har06}. This is another indication that the model presented
here is self-consistent (if only roughly $L_{\rm nuc} \propto L_{\rm
j}$). Moreover, we expect characteristic timescale for the variability
of the emission produced within the HST-1 flaring region $t_{\rm var}
\sim R / c \, \delta \la 1$ yr for $R \la 0.02''$, again in
a rough agreement with the observed one at radio, optical, X-ray, and
$\gamma$-ray frequencies. 

Let us investigate next the expected TeV emission at some later time,
during the synchrotron flare of HST-1, when the nuclear seed photon
energy density in the knot's rest frame has decreased
significantly. Figure 7 shows the expected TeV IC/nuc, IC/star and SSC
luminosities again for $\theta = 20\degr - 30\degr$ and $R= 0.02''$, but
this time with $f_{\rm R} = 40$ mJy, $f_{\rm O} = 200$ $\mu$Jy taken
to illustrate synchrotron continuum of HST-1 flaring region in 2004
\citep{har06}, and $\Gamma_{\rm nuc}^2 L'_{\rm fl} = 10^{45}$ erg
s$^{-1}$ characterizing the quiescence nuclear emission (factor of
$30$ below the high activity epoch considered above). In such a case,
for $\Gamma \sim 2-3$ one expects $L_{\rm IC/star} \sim (1-6) \times
10^{39}$ erg s$^{-1}$ and $L_{\rm SSC} < 10^{39}$ erg s$^{-1}$. Also
$L_{\rm IC/nuc} < L_{\rm IC/star}$ except for the small bulk Lorentz
factors ($\Gamma < 2$) and large jet viewing angle ($\theta \sim
30\degr$). The considered parameters imply now $U'_{\rm B} \sim
10^{-7}$ erg cm$^{-3}$ and $U'_{\rm nuc} \sim 3 \times 10^{-9}$ erg
cm$^{-3}$, i.e. that radiative cooling of the TeV energy electrons is
mainly due to their synchrotron emission. We finally note that the
synchrotron emission of the HST-1 flaring region has increased between
2004 and 2005 by a factor of $2-3$. Thus, in a framework of the
presented model, we expect also the TeV flux due to the IC/star
process to increase in 2005 when compared with the 2004 level.

\section{Summary and Conclusions}

Supermassive black holes present in centers of active galaxies are
known to influence trajectories of nearby stars, and to create in this
way central stellar cusps observed by {\it Hubble}. Here we propose that
the distribution of the hot gas within ellipticals follows closely
distribution of the stars not only in the outer parts of the galaxies,
as observed in a number of such systems, but also in the innermost
parts. If this is the case, then one should expect excess of thermal
pressure (when compared to the pure $\beta$-type profile of the gas
number density) within $\sim 100$ pc from the galactic center. The
resulting small excess in X-ray surface brightness due to free-free radiation
of the hot gas seems to be required to explain some {\it Chandra}
observations. This additional gaseous component can also
result in a stronger confinement of the jets, leading to formation of
stationary reconfinement/reflected shocks within the outflows. We
propose that in the case of the M~87 radio galaxy, HST-1 knot present
at $\sim 100$ pc from the center can be identified with the downstream
region of such a reconfinement/reflected shock. In particular, we
argue that stationary, compact ($R \leq 1$ pc), variable (on the time
scale of, at least, months and years), and overpressured (by a factor
$\ga 10$) HST-1 flaring region is placed immediately downstream of
the point where the converging reconfinement shock reaches the jet
axis (`reconfinement nozzle'). Thereby \emph{some portion} of the hot
relativistic jet decelerates from highly relativistic to mildly
relativistic bulk velocities (from bulk Lorentz factor $\Gamma \sim
10$ down to $\Gamma \sim 2-3$), while other parts of the jet (placed
further away from the jet axis) are expected to decelerate less
strongly due to a larger angle between the upstream bulk velocity vector
and the shock normal. The liberated bulk kinetic energy of the outflow
is transformed at the shock front to the turbulent magnetic field
energy (consistently with the decrease in the degree of linear
polarization observed in HST-1 knot), and, in similar amount (by
assumption), to the ultrarelativistic particles.

Although the reconfinement/reflected shock structure is stationary in
the observer's rest frame, variations and changes in the central engine
lead inevitably to flaring of this part of the outflow, in particular
when the excess particles and photons emitted by the active nucleus in
its high-activity epoch and traveling down the jet arrive after some
time to the reconfinement nozzle. In a framework of this scenario, one
should expect firstly high-energy $\gamma$-ray flare due to
comptonization of the photons from the nuclear outburst, and then,
after some delay depending on the bulk velocity of the nuclear jet,
synchrotron flare due to excess nuclear particles shocked at the
nozzle. This delayed synchrotron flare could be accompanied by the
subsequent inverse-Compton brightening due to upscattering of the
ambient radiation fields by the increased population of the
ultrarelativistic particles. It is tempting to speculate that such a
sequence of events was in fact observed in HST-1 flaring region,
especially as for a realistic set of the jet parameters the evaluated
radiative fluxes are in agreement with the multiwavelength
observations performed between 1998 and 2005. 

\section*{Acknowledgments}

\L .S., M.O., and A.S. were supported by MEiN through the research project 1-P03D-003-29 in years 2005-2008. \L .S.\ acknowledges also the ENIGMA Network through the grant HPRN-CT-2002-00321. M.S. acknowledges partial support from Polish MEiN grant 1-P03D-009-28. A.S. was supported by NASA contract NAS8-39073, and partly by the National Aeronautics and Space Administration through Chandra Award Number GO5-6113X issued by the Chandra X-Ray Observatory Center, which is operated by the Smithsonian Astrophysical Observatory for and on behalf of NASA under contract NAS8-39073. Authors thank Katherine Aldcroft, Daniel E. Harris, and the referee Geoff Bicknell for helpful comments and discussions.

\appendix

\section[]{Reconfinement Shock}

In the rest frame of a shock, the relativistic shock jump conditions can be written as
\begin{equation}
w_- \, \Gamma^2_- \, \beta^2_- + p_- = w_+ \, \Gamma^2_+ \, \beta^2_+ + p_+ \, ,
\end{equation}
\noindent
\begin{equation}
w_- \, \Gamma^2_- \, \beta_- = w_+ \, \Gamma^2_+ \, \beta_+ \, ,
\end{equation}
\noindent
and
\begin{equation}
n_- \, \Gamma_- \, \beta_- = n_+ \, \Gamma_+ \, \beta_+ \, ,
\end{equation}
\noindent
where velocities $\beta_-$ and $\beta_+$ refer to the normal
components of the upstream (`$-$') and downstream (`$+$') bulk
velocity vectors, respectively \citep[see, e.g.,][]{kir99}. Here $w$
is the proper enthalpy of the fluid, $p$ is its isotropic pressure,
and $n$ its the proper number density. Let us consider first the case of
the upstream cold plasma dominated dynamically by the rest energy of
the particles with a mass $m$, with the thermal pressure negligible,
i.e. the enthalpy being approximately equal to the proper
rest-mass energy density $\mu_- \equiv m \, n_- \, c^2$, namely $w_-
\equiv \mu_- + \hat{\gamma} \, p_- / (\hat{\gamma} -1) \approx \mu_-$,
where $\hat{\gamma}$ is the appropriate adiabatic index. One can find
that in such a case
\begin{equation}
p_+ = \mu_- \, \Gamma^2_- \, \beta^2_- \, \left(1- {\beta_+ \over \beta_-}\right) \, .
\end{equation}
\noindent
Now let us consider a supresonic jet which breaks free at some distance
from the central engine, and next experiences reconfinement by the
ambient medium starting from the distance $r_0$. Following
\citet{kom97}, we denote by $\psi$ the angle between the tangent to
the converging reconfinement shock at some given distance $r > r_0$,
and by $\phi$ the angle between the pre-shock jet bulk velocity vector
close to the shock at the same distance $r$. Note, that by the
definition $\tan \psi = - dz / dr$ and $\tan \phi = z(r) / r$, where
$z$ is the distance of the reconfinement shock from the jet axis at
given $r$. We also assume that both angles are small, i.e. $\tan \psi
\approx \psi$ and $\tan \phi \approx \phi$. As the reconfinement shock
is stationary in the observer rest frame, one has
\begin{equation}
\beta_- = \beta_{\rm j} \, \sin \left(\psi + \phi\right) \, ,
\end{equation}
\noindent
where $\beta_{\rm j}$ is the pre-shock jet bulk velocity, and
$\Gamma_{\rm j} \equiv (1 - \beta_{\rm j}^{-2})^{-1/2} = \Gamma_-$ is
the pre-shock bulk Lorentz factor, and, obviously, $\mu_{\rm j} =
\mu_-$. The jet luminosity is $L_{\rm j} = w_{\rm j} \, \Gamma^2_{\rm
j} \, \beta_{\rm j} \, c \, \pi \, R_{\rm j}^2 \approx \mu_{\rm j} \,
\Gamma^2_{\rm j} \, \beta_{\rm j} \, c \, \pi \, r^2 \, \tan^2 \Phi$,
where $R_{\rm j} = r \, \tan \Phi$ is the radius of the free jet and
$\Phi$ is the pre-shock (free) jet opening angle. Taking the external
pressure of the ambient gaseous matter $p_{\rm G}(r) = p_0 \, (r /
r_{\rm B})^{- \eta}$, by means of the condition $p_+(r) = p_{\rm
G}(r)$, one obtains an equation for the distance of the reconfinement
shock from the jet axis
\begin{equation}
{dz \over dr} = {z \over r} - \alpha \, r^{(2 - \eta)/2} \, ,
\end{equation}
\noindent
where
\begin{equation}
\alpha = \left({p_0 \, r_{\rm B}^{\eta} \, c \, \pi \, \tan^2 \Phi \over \zeta_1 \, L_{\rm j} \, \beta_{\rm j}}\right)^{1/2}
\end{equation}
\noindent
and we expressed the term $(1 - \beta_+/\beta_-)$ as a parameter
$\zeta_1$. With the initial condition $z(r_0) = z_0 \equiv r_0 \, \tan
\Phi$, the solution to the above equation,
\begin{equation}
z(r) = r \, \tan \Phi - {2 \, \alpha \over 2 - \eta} \, r \, \left(r^{(2 - \eta)/2} - r_0^{(2 - \eta)/2}\right) \, ,
\end{equation}
\noindent
implies that the reconfinement shock reaches the jet axis at the distance
\begin{equation}
r_{\rm cr} \approx \left[ {(2 - \eta)^2 \, \zeta_1 \over 4} \, {L_{\rm j} \over p_0 \, r_{\rm B}^{\eta} \, c \, \pi} \right]^{1/(2 - \eta)} \, .
\end{equation}
\noindent
Note, that at $r_0$ one has
\begin{equation}
\tan \Phi = {2 \over (\hat{\gamma} - 1) \, \mathcal{M}_{\rm j}} \equiv {2 \, \Gamma_{\rm s, \, j} \, \beta_{\rm s, \, j} \over (\hat{\gamma} - 1) \, \Gamma_{\rm j} \, \beta_{\rm j}}  \, ,
\end{equation}
\noindent
where $\mathcal{M}_{\rm j}$ is the relativistic Mach number of a free
jet, and $\beta_{\rm s, \, j} \equiv (1 - \Gamma_{\rm s, \,
j}^{-2})^{1/2}$ is the jet sound speed in $c$ units. For a cold jet
matter considered here $\Gamma_{\rm s, \, j} \, \beta_{\rm s, \, j}
\approx \beta_{\rm s, \, j} = (\hat{\gamma} \, p_{\rm j} / \mu_{\rm
j})^{1/2}$. Since at $r_0$ jet pressure equals external gas pressure,
$p_{\rm j}(r_0) = p_{\rm G}(r_0)$, one can therefore find that
\begin{equation}
L_{\rm j} = {4 \hat{\gamma} \over (\hat{\gamma} - 1)^2} \, c \, \pi \, p_0 \, r_{\rm B}^{\eta} \, r_0^{2 - \eta} \, .
\end{equation}
\noindent
This, together with the equation for $r_{\rm cr}$, gives the condition
\begin{equation}
{r_{\rm cr} \over r_0} = \left[{(2 - \eta)^2 \, \zeta_1 \, \hat{\gamma} \over (\hat{\gamma} - 1)^2}\right]^{1/(2 - \eta)} \, .
\end{equation}
\noindent
With $\hat{\gamma} = 5/3$ one obtains $r_{\rm cr} / r_0 =  \left[2.625 \, (2 - \eta)^2 \right]^{1/(2 - \eta)}$ for $\zeta_1 \approx 0.7$ \citep[see][]{kom97}.
 
Now let us consider an analogous case as before, but with a jet matter
described by an ultrarelativistic equation of state, $w = 4 \, p$
(i.e., with $\hat{\gamma} = 4/3$). In this case the upstream pressure
cannot be neglected anymore, and by combining equations A1-A2 one
obtains for a relativistic jet
\begin{equation}
p_+ = p_- \left[ 4 \, \Gamma^2_- \, \beta^2_- \, \left(1- {\beta_+ \over \beta_-}\right) + 1 \right] \approx p_- \, 4 \, \Gamma^2_- \, \beta^2_- \, \zeta_2 \, ,
\end{equation}
\noindent
where $\zeta_2 \equiv 1- (\beta_+ / \beta_-) = 1 - (1/3 \, \beta_-^2) \approx 0.65$. With the appropriate expression for the jet kinetic luminosity, $L_{\rm j} = 4 \, p_{\rm j} \, \Gamma^2_{\rm j} \, \beta_{\rm j} \, c \, \pi \, r^2 \, \tan^2 \Phi$, one obtains again
\begin{equation}
r_{\rm cr} \approx \left[ {(2 - \eta)^2 \, \zeta_2 \over 4} \, {L_{\rm j} \over p_0 \, r_{\rm B}^{\eta} \, c \, \pi} \right]^{1/(2 - \eta)} \, .
\end{equation}
\noindent
In addition, in the case of the ultrarelativistic equation of state the sound speed is $\beta_{\rm c, \, j} = 1 / \sqrt{3}$, and hence by means of expression A10 in a form
\begin{equation}
\Gamma_{\rm j} \, \beta_{\rm j} = {3 \, \sqrt{2} \over \tan \Phi} \, ,
\end{equation}
\noindent 
one obtains condition $r_{\rm cr} / r_0 = \left[18 \, (2 - \eta)^2 \, \zeta_2 \right]^{1/(2 - \eta)} \approx \left[11.7 \, (2 - \eta)^2 \right]^{1/(2 - \eta)}$.

\section[]{Radiative Formulae}

For a given radio flux $f_{\rm R}$ as measured at some observed radio
frequency $\nu_{\rm R}$, and for the observed emission region size
$R$, the intensity of the equipartition magnetic field evaluated ignoring
relativistic correction is
\begin{equation}
B_{\rm eq, \, \delta=1} \propto \left[\nu_{\rm R}^{\alpha} f_{\rm R} \, \left( \nu_{\rm min}^{-\alpha+1/2} - \nu_{\rm max}^{-\alpha+1/2} \right) \, R^{-3} \right]^{2/7} \, ,
\end{equation}
\noindent
where $\nu_{\rm min}$ and $\nu_{\rm max}$ are the minimum and maximum
frequencies of the synchrotron continuum, assumed to be a simple
power-law characterized by a spectral index $\alpha$ \citep[see,
e.g.,][]{lon94}. As discussed by \citet{sta03}, relativistic
corrections give $B_{\rm eq} = B_{\rm eq, \, \delta = 1} \,
\delta^{-5/7}$. Hence, taking $\nu_{\rm max} \gg \nu_{\rm min} \sim
\nu_{\rm R}$ and $\alpha > 0.5$, one obtains
\begin{equation}
B_{\rm eq} \propto \nu_{\rm R}^{1/7} \, f_{\rm R}^{2/7} \, R^{-6/7} \, \delta^{-5/7}
\end{equation}
\noindent
\citep[see in this context][]{kat05}. This gives the comoving minimum magnetic field energy density $U'_{\rm B} \propto B_{\rm eq}^2 \propto f_{\rm R}^{4/7} \, R^{-12/7} \, \delta^{-10/7}$.

As discussed in \citet{sta03}, the observed inverse-Compton
luminosities (produced in the Thomson regime) can be simply evaluated
as
\begin{equation}
L_{\rm IC/seed} \sim f_{\rm \pm, \, iso} \, {U'_{\rm seed} \over U'_{\rm B}} \, L_{\rm syn} \, ,
\end{equation}
\noindent
where $L_{\rm syn}$ is the observed synchrotron luminosity, $U'_{\rm
seed}$ is the comoving energy densities of the seed photons, while
$f_{\rm \pm, \, iso} = f_{\rm \pm, \, iso}(\Gamma, \theta)$ is the
function of the kinematic jet parameters arising due to possible
anisotropy of the seed photon fields in the jet rest frame. In section
5, the observed bolometric synchrotron luminosity is approximated by
the optical one, $L_{\rm syn} \propto \nu_{\rm O} f_{\rm O}$. 
In section 5 we also assumed that at every moment (i.e., for a given
synchrotron flux), the emission region is in the equipartition regarding
energies of the radiating electrons and the magnetic field.

In the case of the synchrotron self-Compton emission, $f_{\rm iso} = 1$ and $U'_{\rm syn} \propto f_{\rm O} \, R^{-2} \, \delta^{-3}$, leading to
\begin{equation}
L_{\rm SSC} \propto f_{\rm R}^{-4/7} \, f_{\rm O}^{2} \, R^{-2/7} \,
\delta^{-11/7} \, .
\end{equation}
\noindent
If the comptonisation of the starlight emission is considered, $f_{+} \sim (\delta / \Gamma)^2$ and $U'_{\rm star} \propto \Gamma^2$. Hence,
\begin{equation}
L_{\rm IC/star} \propto f_{\rm R}^{-4/7} \, f_{\rm O} \, R^{12/7} \, \delta^{24/7} \, .
\end{equation}
\noindent
Finally, for the comptonisation of the nuclear emission illuminating
the jet from behind, we have  $f_{-} \sim \delta^2 \, \Gamma^2 \, (1 -
\cos \theta)^2$ and $U'_{\rm nuc} \propto (L'_{\rm fl} \Gamma_{\rm
nuc}^2) \, \Gamma^{-2} \, (\sin \theta)^{2}$, where the factor $(\sin
\theta)^2$ is due to deprojecting the observed distance of the
emission region.
This leads to
\begin{equation}
L_{\rm IC/nuc} \propto f_{\rm R}^{-4/7} \, f_{\rm O} \, R^{12/7} \, (L'_{\rm fl} \Gamma_{\rm nuc}^2) \, \delta^{24/7} \, (\sin \theta)^{2} \, (1 - \cos \theta)^2 \, .
\end{equation}
\noindent

The approximations derived above allow us to estimate in a simple way the
expected $\gamma$-ray fluxes for a given $f_{\rm R}$, $f_{\rm O}$,
$R$, and the kinematic parameters of the jet.

\label{lastpage}

\end{document}